\documentclass[aps,pra,twocolumn,groupedaddress,floatfix,showpacs]{revtex4}

\usepackage{graphicx}
\usepackage[latin1]{inputenc}
\usepackage{amsmath}
\usepackage{amssymb}
\usepackage{color}

\newcommand{\bra}[1]{\ensuremath{\langle{#1}\vert}}
\newcommand{\ket}[1]{\ensuremath{\vert{#1}\rangle}}

\newcommand{\mean}[1]{\ensuremath{\langle{#1}\rangle}}
\newcommand{\abs}[1]{\ensuremath{\vert{#1}\vert}}

\begin{document}

\title{Superradiant Decay and Dipole-Dipole Interaction of Distant Atoms\\ in a Two-Way Cascaded Cavity QED System}
\author{Steffen Zeeb$^{1}$, Changsuk Noh$^{2}$, A.~S. Parkins$^{3}$\footnote{Corresponding author: s.parkins@auckland.ac.nz}, and H.~J. Carmichael$^{3}$}
\address{$^1$ Institute of Theoretical Physics, University of W\"urzburg, 97074 W\"urzburg, Germany}
\address{$^2$ Centre for Quantum Technologies, National University of Singapore, 3 Science Drive 2, Singapore 117543}
\address{$^3$ The Dodd-Walls Centre for Photonic and Quantum Technologies, and Department of Physics, University of Auckland, Private Bag 92019, Auckland, New Zealand}

\date{\today}

\begin{abstract}
We investigate a two-way cascaded cavity QED system consisting of microtoroidal resonators coupled through an optical fiber. Each microtoroidal cavity supports two counter-propagating whispering-gallery modes coupled to single atoms through their evanescent fields. We focus on a pair of atom-microtoroid systems and compute the spectrum of spontaneous emission into the fiber with one atom initially excited. Explicit results are presented for strong-coupling and bad-cavity regimes, where the latter allows the effective atom-atom interaction to be controlled through the atom-cavity coupling and detuning: the atoms exhibit either collective spontaneous emission with no dipole-dipole interaction or a (coherent) dipole-dipole interaction and independent (single-atom) emission. This capacity for switching the character of the interaction is a feature of bi-directional coupling and connects our two-way cascaded system to work on one-dimensional waveguides. Building upon our bad-cavity results, we generalize to many atom-microtoroid systems coupled through an optical fiber.
\end{abstract}


\maketitle

\section{Introduction}

With a view to realizing distributed quantum networks in which atomic or solid state qubits are coupled through traveling light fields over substantial distances, a number of new architectures for cavity quantum electrodynamics (CQED) are proposed \cite{Vahala04,Kimble08}. These include monolithic microtoroids and microdisks, whose whispering gallery modes may couple to free atoms via their evanescent fields \cite{Aoki06,Dayan08,Aoki09,Alton10,Junge13,OShea13,Shomroni14}, or quantum dots embedded in the resonators themselves \cite{Kiraz01,Peter05,Srinivasan07a,Srinivasan08}. Highly efficient input and output is achieved by overlapping the evanescent fields of tapered optical fibers with the resonator modes. Fibers are also the natural choice for linking resonators, i.e., as a channel of communication between different nodes of a network of such microresonator CQED systems.

Quite apart from applications to quantum information science, such systems offer an intriguing new environment in which to study basic CQED phenomena, e.g.,  modified atomic spontaneous emission, where distant atoms interact through the resonator and fiber fields. New features compared with conventional single-mode CQED include counterpropagating resonator and fiber modes, and the open-system nature of the fiber-mediated coupling, where the fiber acts as a common, broadband reservoir to the modes of the microtoroids. A natural modeling of this coupling is provided within the framework of cascaded open quantum systems \cite{Gardiner93,Carmichael93}. Starting from such a model, we focus for the most part in this paper on a pair of two-way cascaded atom-microtoroid systems \cite{Zeeb09}---the many-site extension is briefly considered---and compute the spectrum of spontaneous emission when one atom is initially excited. After deriving general expressions for all system outputs, including side emission from the atoms and scattering in the microtoroids, we concentrate on the readily accessible spontaneous emission through the counter-propagating fiber modes.

Our model, most generally, describes a rich dynamic; the initial excitation is shared between two atoms and four microtoroid modes. We discuss two limiting cases in detail: the strong-coupling regime, with the atom-cavity coupling constants much larger than the system decay rates, and the bad-cavity (so-called `Purcell') regime, where the cavity decay rates are sufficiently large that an adiabatic elimination of the microtoroid fields may be performed. In the strong coupling regime, depending on the relative coupling phase,  we show that either a vacuum Rabi oscillation develops between delocalized---shared between the two atom-microtoroid systems---atomic and photonic excitations, or localized atom-photon dressed states couple through the cavity decay into the fiber.

In the bad-cavity regime, after adiabatic elimination, again depending on relative coupling phase (alternatively, the atom-cavity detuning), the dynamics simplify to yield strongly contrasting forms of the effective---mediated by the microtoroid and fiber fields---atom-atom interaction: the atoms either exhibit collective spontaneous emission through the fiber with no dipole-dipole interaction, or a fiber-mediated dipole-dipole interaction with cavity-enhanced {\em single-atom} spontaneous emission. We recover behavior seen in the interaction of distant atoms through a one-dimensional waveguide
\cite{LeKien05,Dzsotjan10,Chang12,Lalumiere13,vanLoo13,Gonzalez-Tudela13}.

We present our cascaded open quantum system model for a pair of atom-microtoroid systems in Sec.~\ref{sec:model}, and derive general expressions for the spontaneous emission spectra at its various outputs in Sec.~\ref{sec:spectra}. We then discuss the contrasting behaviors in the strong-coupling regime, Sec.~\ref{sec:strong-coupling}, and the bad-cavity regime, Sec.~\ref{sec:bad-cavity}. In Sec.~\ref{sec:bad-cavity} we derive a master equation for the atoms alone, which provides a compact and transparent description of the atomic dynamics, and connects our two-way cascaded system, in the bad-cavity regime, to a one-dimensional waveguide. The master equation is generalized to many two-way cascaded atom-microtoroid systems in Sec.~\ref{sec:many_sites}, where similar---but not the same---contrasting behaviors as a function of coupling phase are discussed. Spectra for $N=3$ two-way cascaded atom-microtoroid systems illutrate these results. Sec.~\ref{sec:conclusion} provides a summary and conclusions.

\section{Theoretical Model}
\label{sec:model}

\begin{figure}[h]
\begin{center}
\includegraphics[width=8cm]{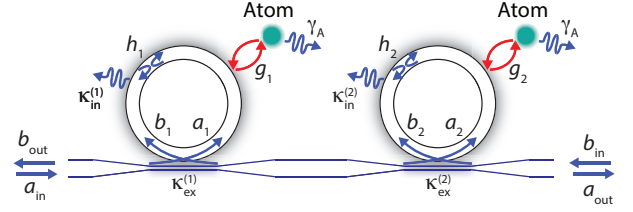}
\caption{Schematic of the cascaded atom-microtoroid system. Counter-propagating microtoroid modes couple through their evanescent fields to two two-state atoms and the input-output fields of an optical fiber.}
\label{fig:fig1}
\end{center}
\end{figure}

We begin by investigating a system of two microtoroids coupled through an optical fiber, as shown in Fig.~\ref{fig:fig1}. The microtoroids are distinguished by a label $i=1,2$. Toroid $i$ supports a pair of counter-propagating whispering gallery modes (WGM's) of frequency $\omega_{\mathrm{C}_i}$, with each coupled to the fiber through evanescent fields at loss rate $2\kappa_\mathrm{ex}^{(i)}$. Light is lost out the sides of the microtoroids---e.g., due to scattering from imperfections---at an intrinsic loss rate $2\kappa_\mathrm{in}^{(i)}$. A pair of identical two-level atoms, transition frequency $\omega_\mathrm{A}$ and spontaneous emission rate $\gamma_{\rm A}$, couple to the WGM's through evanescent fields with coupling constants $g_i$, $i=1,2$.

The system is modeled as a two-way cascaded system. We introduce creation and annihilation operators $a_i$, $b_i$ (clockwise, counterclockwise) and $a_i^\dagger$, $b_i^\dagger$ for the WGM's, and raising and lowering operators $\sigma_i^\pm$ for the two-level atoms. Then with Lindblad operator $D[{\cal O}]\cdot =2{\cal O}\cdot{\cal O}^\dag - {\cal O}^\dag{\cal O}\cdot - \cdot{\cal O}^\dag{\cal O}$,  the master equation, in a frame rotating at frequency $\omega_\mathrm{A}$, is written as \cite{Gardiner93,Carmichael93,Carmichael07}
\begin{widetext}
\begin{align}
\dot{\rho}=&-i\left[ H_1+H_2+H_{12},\rho\right]+\frac12\mkern-2mu\left(D[J_a]\rho+D[J_b]\rho\right)\notag\\
&+\kappa_\mathrm{in}^{(1)}\mkern-2mu\left(D[a_1]\rho+D[b_1]\rho\right)+\kappa_\mathrm{in}^{(2)}\mkern-2mu\left(D[a_2]\rho+D[b_2]\rho\right)
+\frac{\gamma_{\rm A}}{2}\mkern-2mu\left(D[\sigma_1^-]\rho+D[\sigma_2^-]\rho\right),
\label{eqn:master_equation1}
\end{align}
where $\rho$ is the reduced density operator of the microtoroid-plus-atoms system, and (setting $\hbar=1$) the Hamiltonian is defined by
\begin{subequations}
\begin{align}
H_1&=\Delta_1(a_1^\dagger a_1+b_1^\dagger b_1)+(h_1a_1^\dagger b_1+h_1^* b_1^\dagger a_1)+(g_1^*a_1^\dagger\sigma_1^-+g_1\sigma_1^+a_1 )
+(g_1 b_1^\dagger\sigma_1^-+g_1^*\sigma_1^+ b_1),
\label{eqn:hamiltonian}\\
\noalign{\vskip4pt}
H_2&=\Delta_2(a_2^\dagger a_2+b_2^\dagger b_2)+(h_2 a_2^\dagger b_2+h_2^* b_2^\dagger a_2)+(g_2^* a_2^\dagger\sigma_2^-+g_2\sigma_2^+a_2)
+(g_2 b_2^\dagger\sigma_2^-+g_2^*\sigma_2^+b_2),
\end{align}
with $\Delta_i=\omega_{\mathrm{C}_i}-\omega_\mathrm{A}$, and
\begin{equation}
H_{12}=i\sqrt{ \kappa_\mathrm{ex}^{(1)}\kappa_\mathrm{ex}^{(2)}}(e^{-i\phi_a}a_1^\dagger a_2-e^{i\phi_a}a_2^\dagger a_1)
+i\sqrt{\kappa_\mathrm{ex}^{(1)}\kappa_\mathrm{ex}^{(2)}}(e^{-i \phi_b} b_2^\dagger b_1-e^{i \phi_b}b_1^\dagger b_2).
\end{equation}
\end{subequations}
\end{widetext}
$H_1$ describes the first atom-microtoroid system, $H_2$ the second, and $H_{12}$ the coupling between them via the fiber. Within each microtoroid, the counter-propagating modes can couple directly with strength $h_i$, and $\phi_{a,b}$ are phases that account for the propagation distance between the microtoroids. The operators
\begin{subequations}
\begin{align}
J_a&=\sqrt{2\kappa_\mathrm{ex}^{(1)}}a_1+e^{-i\phi_a}\sqrt{2\kappa_\mathrm{ex}^{(2)}}a_2,\\
\noalign{\vskip2pt}
J_b&=\sqrt{2\kappa_\mathrm{ex}^{(2)}}b_2+e^{-i\phi_b}\sqrt{2 \kappa_\mathrm{ex}^{(1)}}b_1,
\end{align}
\end{subequations}
describe a two-way cascaded coupling between the modes of the two microtoroids mediated by the fiber, which is regarded to provide a common (broadband) reservoir for these modes. Note that with the propagation directions coupled---through scattering between counterpropagating modes ($h_{1,2}$) and from the atoms ($g_{1,2}$)---the model as written is an approximation, in so far as it neglects retardation associated with the time of flight between the microtoroids. The approximation is valid provided the flight time is much smaller than the timescale on which the system state undergoes significant change \cite{Carmichael07}, which in practice puts a limit on the length of fiber connecting the microtoroids.

To summarize, the model incorporates three different types of damping: loss from the WGM's into the fiber at rate $2\kappa_\mathrm{ex}$, intrinsic scattering loss from the WGM's at rate $2\kappa_\mathrm{in}$, and atomic spontaneous emission to free space at rate $\gamma_{\rm A}$. In addition, the counterpropagating modes of each microtoroid are directly coupled ($h_{1,2}$) and each atom couples at equal strength to both modes of its respective microtoroid ($g_{1,2}$).

\section{Spectrum of spontaneous emission}
\label{sec:spectra}

\subsection{Quantum trajectory unraveling}
\label{sec:qtunravel}
In this section we wish to calculate the spectrum of emitted photons starting from an initial state with one atom excited and the other in the ground state. We use the method introduced in Chap.~13 of \cite{Carmichael07}, which writes the master equation as $\dot{\rho} = \mathcal{L} \rho$ and separates $\mathcal{L}$ into two parts: the first part acts only on the one-quantum subspace, and the second is the generator of transitions out of this subspace to the ground state. Effectively we implement a quantum trajectory unraveling and compute the spectrum from the response of a frequency selective detector placed in the output field (Sec.~19.3.2 of \cite{Carmichael07}).

With $\mathcal{L}$ decomposed in this way, the master equation is written as
\begin{equation}
\dot\rho=(\mathcal{C}+\mathcal{D})\rho,
\label{eqn:dotrho}
\end{equation}
with superoperator $\mathcal{C}$ acting only in the one-quantum subspace:
\begin{align}
\mathcal{C}\,\cdot=&-i[H_1+H_2+H_{12},\cdot\,]-\frac12[J_a^\dagger J_a+J_b^\dagger J_b,\cdot\,]_+\notag\\
\noalign{\vskip2pt}
&-\kappa_\mathrm{in}^{(1)}[a_1^\dagger a_1+b_1^\dagger b_1,\cdot\,]_+-\kappa_\mathrm{in}^{(2)}[a_2^\dagger a_2+b_2^\dagger b_2,\cdot\,]_+\notag\\
\noalign{\vskip6pt}
&-\frac{\gamma_{\rm A}}{2}[\sigma_1^+\sigma_1^-+\sigma_2^+\sigma_2^-,\cdot\,]_+,
\label{eqn:defC}
\end{align}
with $[\,\cdot\,,\, \cdot \,]_+$ the anti-commutator, and superoperator $\mathcal{D}$ generating transitions to the ground state:
\begin{align}
\mathcal{D}\,\cdot=&\,\,J_a\cdot J_a^\dagger+J_b \cdot J_b^\dagger\notag\\
\noalign{\vskip2pt}
&+2\kappa_\mathrm{in}^{(1)}(a_1\cdot\, a_1^\dagger+b_1\cdot\, b_1^\dagger)+2\kappa_\mathrm{in}^{(2)}(a_2\cdot\, a_2^\dagger+b_2 \cdot\, b_2^\dagger)\notag\\
\noalign{\vskip2pt}
&+\gamma_{\rm A}(\sigma_1^-\cdot\,\sigma_1^++\sigma_2^-\cdot\,\sigma_2^+).
\label{eqn_defD}
\end{align}
The initial density operator with the first atom excited is written $\rho(0)=\rho_A(0)\rho_{ab}(0)$, with atomic state
\begin{equation}
\rho_A(0)=(\ket{\rm e}\bra{\rm e})_{A_1}(\ket{\rm g}\bra{\rm g})_{A_2},
\end{equation}
and $\rho_{ab}(0)$ the vacuum state of the field.

Now for times $t>0$, a photon has either been emitted, probability $P_{\mathrm{spon}}(t)$, in which case the system is in the ground state,
\begin{equation}
\ket{G}=\ket{\rm g}_{A_1}\ket{\rm g}_{A_2}\ket{0}_{a_1}\ket{0}_{b_1}\ket{0}_{a_2}\ket{0}_{b_2},
\end{equation}
or no photon has been emitted and the system is in a pure one-quantum state:
\begin{align}
\ket{\bar{\psi}(t)}=&\,\big[\xi_1(t)\sigma_1^++\alpha_1(t)a_1^\dagger+\beta_1(t)b_1^\dagger\notag\\
\noalign{\vskip2pt}
&+\xi_2(t)\sigma_2^++\alpha_2t)a_2^\dagger+\beta_2t)b_2^\dagger\big]\ket{G},
\label{eqn:state}
\end{align}
where $\xi_1(t)$ is the probability amplitude for the first atom to be still excited, $\alpha_1(t)$ is the probability amplitude for the transfer of the excitation to cavity mode $a_1$, and so on as indicated by the defined photon creation and atomic raising operators. From this the density operator, at all times, is a sum of two parts: $\rho(t)=\rho_0(t)+\rho_1(t)$, with
\begin{equation}
\rho_0(t)=P_{\mathrm{spon}}(t)\ket{G}\bra{G},
\label{eqn:rho0}
\end{equation}
and
\begin{equation}
\rho_1(t)=\ket{\bar{\psi}(t)}\bra{\bar{\psi}(t)},
\label{eqn:rho1}
\end{equation}
where the norm of $\ket{\bar\psi(t)}$ is $1-P_{\mathrm{spon}}(t)$, i.e., it is the null-record probability, the probability that no photon has been emitted.
As $\mathcal{C}$ acts only in the one-quantum subspace (the between-jump evolution), and $\mathcal{D}$ takes $\ket{\bar\psi(t)}$ to $\ket{G}$ (the quantum jump), we find from Eq.~(\ref{eqn:dotrho}):
\begin{align}
\dot{\rho}_1(t)&=\mathcal{C}\rho_1,
\label{eqn:dotrho1}\\
\noalign{\vskip2pt}
\dot{\rho}_0(t)&=\mathcal{D}\rho_1.
\label{eqn:dotrho0}
\end{align}
The equation of motion for the emission probability follows from Eq.~(\ref{eqn:dotrho0}),
\begin{align}
\frac{dP_{\mathrm{spon}}}{dt}&=\sum_{i}2\kappa_i\left[\abs{\alpha_i(t)}^2+\abs{\beta_i(t)}^2\right]+\gamma_{\rm A}\abs{\xi_i(t)}^2,\notag\\
\noalign{\vskip-10pt}
\end{align}
where $\kappa_i=\kappa_\mathrm{in}^{(i)} + \kappa_\mathrm{ex}^{(i)}$, and Eq.~(\ref{eqn:dotrho1}) yields the non-unitary Schr\"odinger equation
\begin{equation}
\frac{d\ket{\bar{\psi}(t)}}{dt}=-i H\ket{\bar{\psi}(t)} \, ,
\label{eqn:non_unitary_evolution}
\end{equation}
with non-Hermitian Hamiltonian
\begin{align}
H =&\,\, H_1+H_2+H_{12} - i \frac{1}{2} \big(J_a^\dagger J_a + J_b^\dagger J_b\big)\notag\\
&-i\kappa_\mathrm{in}^{(1)}(a_1^\dagger a_1 +  b_1^\dagger b_1) -i\kappa_\mathrm{in}^{(2)}( a_2^\dagger a_2 +b_2^\dagger b_2)\notag\\
\noalign{\vskip4pt}
&-i \frac{\gamma_{\rm A}}2(\sigma_1^+ \sigma_1^- +\sigma_2^+ \sigma_2^-).
\end{align}
The nonunitary Schr\"odinger equation leads to a system of coupled differential equations satisfied by the probability amplitudes:
\begin{subequations}
\begin{align}
\dot\xi_1=&-\frac{\gamma_{\rm A}}{2}\xi_1-ig_1\alpha_1-ig_1^*\beta_1,
\label{eqn:eqns_of_motiona}\\
\noalign{\vskip3pt}
\dot\alpha_1=&-\left(\kappa_1+i\Delta_1\right)\alpha_1-ig_1^*\xi_1-ih_1\beta_1,\\
\noalign{\vskip3pt}
\dot\beta_1=&-\left(\kappa_1+i\Delta_1\right)\beta_1\notag\\
&-ig_1\xi_1-ih_1^*\alpha_1-2\sqrt{\kappa_\mathrm{ex}^{(1)}\kappa_\mathrm{ex}^{(2)}}e^{i \phi_b}\beta_2,\\
\noalign{\vskip3pt}
\dot\xi_2=&-\frac{\gamma_{\rm A}}{2}\xi_2-ig_2\alpha_2-ig_2^*\beta_2,
\label{eqn:eqns_of_motiond}\\
\noalign{\vskip3pt}
\dot\alpha_2=&-\left(\kappa_2+i\Delta_2\right)\alpha_2\notag\\
&-ig_2^*\xi_2-ih_2\beta_2-2\sqrt{\kappa_\mathrm{ex}^{(1)}\kappa_\mathrm{ex}^{(2)}}e^{i \phi_a}\alpha_1,\\
\noalign{\vskip3pt}
\dot\beta_2=&-\left(\kappa_2+i\Delta_2\right)\beta_2-ig_2\xi_2-ih_2^*\alpha_2,
\label{eqn:eqns_of_motionf}
\end{align}
\end{subequations}
Emission spectra for the various outputs can be expressed in terms of the Laplace transform of these probability amplitudes.

\subsection{Emission spectra}
Spontaneous emission is a non-stationary process. Its spectrum can be computed from a double integration over a two-time correlation function, a formula  derived from the response of a tunable oscillator driven by the emitted light \cite{Glauber65}, e.g., a tunable filter cavity cascaded with the source of the spontaneous emission (Sec.~19.3.2 of \cite{Carmichael07}). In the present case, the spontaneous emission is divided between several output channels; we treat them one by one.

\subsubsection{Side emission}
Emission out the sides of the microtoroids---to follow the nomenclature of cavity QED---consists first of atomic spontaneous emission to all modes other than $a_i$ and $b_i$, i.e., atomic emission into free space. The atomic emission spectra are ($i=1,2$)
\begin{equation}
T_{\mathrm{side,}\sigma}^{(i)}(\omega)=\frac{\gamma_{\rm A}}{2\pi}\int_0^{\infty}\mkern-3mu dt\int_0^{\infty}\mkern-3mu dt^\prime e^{-i\omega(t-t^\prime)} \mean{\sigma^+_i(t)\sigma^-_i(t')},
\label{eqn:spectrum_def}
\end{equation}
where, from the quantum regression formula, with $t^\prime\geq t$,
\begin{equation}
\mean{\sigma^+_i(t)\sigma^-_i(t^\prime)}=\mathrm{Tr}\left\{\sigma^-_ie^{\mathcal{(C}+\mathcal{D})(t^\prime-t)}\left[\rho(t)\sigma^+_i\right] \right\}.
\label{eqn:quantum_regression}
\end{equation}
More generally, the upper limit of the integral in Eq.~(\ref{eqn:spectrum_def}) replaced by a finite time $T$ and a spectral filter bandwidth, $\Gamma>0$, may be introduced. Such time-dependent spectra for one-way cascaded microtoroids are calculated in \cite{Mirza13}.

Using the unraveling of Sec.~\ref{sec:qtunravel}, Eqs.~(\ref{eqn:spectrum_def}) and (\ref{eqn:quantum_regression}) lead to a spectrum expressed as the squared modulus of the Laplace transform of $\xi_i(t)$. We may first write
\begin{equation}
\rho(t) \sigma^+_i=\xi_i^*(t)\ket{\bar{\psi}(t)} \bra{G}.
\end{equation}
Noting then that $(\mathcal{C}+\mathcal{D})\left[\ket{\bar{\psi}(t)}\bra{G}\right]=-i H\ket{\bar{\psi}(t)}\bra{G}$, we find
\begin{align}
e^{\mathcal{(C}+\mathcal{D})(t^\prime-t)}\left[\ket{\bar{\psi}(t)}\bra{G}\right]&=e^{-i H(t^\prime-t)}\ket{\bar{\psi}(t)}\bra{G}\notag\\
&=\ket{\bar{\psi}(t^\prime)}\bra{G},
\end{align}
and hence, from Eq.~(\ref{eqn:quantum_regression}),
\begin{align}
\mean{\sigma^+_i(t)\sigma^-_i(t^\prime)}&=\xi_i^*(t)\mathrm{tr}\left[\sigma_i^-\ket{\bar{\psi}(t^\prime)}\bra{G}\right]\notag\\
\noalign{\vskip2pt}
&=\xi_i^*(t) \xi_i(t'),\qquad t^\prime\geq t.
\end{align}
With $\mean{\sigma^+_i(t)\sigma^-_i(t^\prime)}=\mean{\sigma^+_i(t^\prime)\sigma^-_i(t)}^*$, the result holds for $t^\prime<t$ as well. We arrive, thus, at the atomic spontaneous emission spectra ($i=1,2$):
\begin{equation}
T_{\mathrm{side,}\sigma}^{(i)}(\omega)=\frac{\gamma_{\rm A}}{2\pi}\left| \tilde{\xi_i}(\omega)\right|^2,
\label{eqn:spectrum_atoms}
\end{equation}
where $\tilde{\xi_i}(\omega)$ denotes the Laplace transform evaluated at $s=-i\omega$:
\begin{equation}
\tilde{\xi_i}(\omega)=\int_0^{\infty}\mkern-3mudt\:e^{i\omega t}\xi_i(t).
\end{equation}

Spontaneous emission spectra for the four microtoroid scattering outputs might be defined and calculated in the same way. Most generally, however, the side emission might not be resolved and assigned to individual sources. The most natural generalization, built upon the method above, is therefore
\begin{equation}
T_{\mathrm{side}}(\omega)=\left|\sum_i\left[E_i\tilde\xi_i(\omega)+A_i\tilde\alpha_i(\omega)+B_i\tilde\beta_i(\omega)\right]\right|^2,
\label{eqn:spectrum_side}
\end{equation}
with $E_i$, $A_i$, and $B_i$ complex constants determined by the way the side-light is collected and hence interferes.

\subsubsection{Fiber emission}

The spectra of primary interest relate to emission from the microtoroid WGM's into the fiber outputs. There are two fiber outputs, one, output $a$, traveling from left to right, and the other, output $b$, traveling from right to left. In parallel with Eq.~(\ref{eqn:spectrum_def}), the two fiber emission spectra are defined by
\begin{widetext}
\begin{subequations}
\begin{align}
T_{\mathrm{fiber},a}(\omega)&=\frac{1}{2\pi}\int_0^{\infty}\mkern-3mudt\int_0^{\infty}\mkern-3mudt^\prime e^{-i\omega(t-t^\prime)}
\mean{a_{\mathrm{out}}^\dagger(t)a_{\mathrm{out}}(t^\prime)},\\
\noalign{\vskip4pt}
T_{\mathrm{fiber},b}(\omega)&=\frac{1}{2\pi}\int_0^{\infty}\mkern-3mudt\int_0^{\infty}\mkern-3mudt^\prime e^{-i\omega(t-t^\prime)}
\mean{b_{\mathrm{out}}^\dagger(t) b_{\mathrm{out}}(t^\prime)},
\end{align}
\end{subequations}
where $a_\mathrm{{out}}$ and $b_\mathrm{{out}}$ are annihilation operators for output fields traveling in opposite directions \cite{CollettGardiner84_85}:
\begin{subequations}
\begin{align}
a_{\mathrm{out}}(t)&=- a_{\mathrm{in}}(t) + J_a(t),\\
b_{\mathrm{out}}(t)&=- b_{\mathrm{in}}(t) + J_b(t),
\end{align}
\end{subequations}
where $a_{\mathrm{in}}$ and $b_{\mathrm{in}}$ are the corresponding inputs---in our case vacuum fields. By following the steps leading to Eqs.~(\ref{eqn:spectrum_atoms}) and (\ref{eqn:spectrum_side}), the spectra are calculated as
\begin{subequations}
\begin{align}
T_{\mathrm{fiber},a}(\omega)&=\frac{1}{\pi}\left|\sqrt{\kappa_\mathrm{ex}^{(1)}}\tilde{\alpha}_1(\omega)+e^{-i\phi_a}\sqrt{\kappa_\mathrm{ex}^{(2)}}
\tilde{\alpha}_2(\omega)\right|^2,
\label{eqn:spectrum_fibera}\\
\noalign{\vskip4pt}
T_{\mathrm{fiber},b}(\omega)&=\frac{1}{\pi}\left|\sqrt{\kappa_\mathrm{ex}^{(1)}}\tilde{\beta}_1(\omega)+e^{i\phi_b}\sqrt{\kappa_\mathrm{ex}^{(2)}} \tilde{\beta}_2(\omega)\right|^2.
\label{eqn:spectrum_fiberb}		
\end{align}
\end{subequations}
\end{widetext}

\section{Strong-coupling regime}
\label{sec:strong-coupling}

We focus in this section and the next on two regimes of special interest in what, more generally, is a very large parameter space. We first consider strong coupling, i.e., conditions where the dipole coupling constants are much larger than the dissipative rates, where $g_i\gg\{\kappa_i,\gamma_{\rm A}\}$.
To simplify matters, we assume the two microtoroid systems identical, with
\begin{equation}
\Delta_1=\Delta_2=\Delta,\qquad \kappa_1=\kappa_2=\kappa,
\label{eqn:simplification1}
\end{equation}
and also that intrinsic losses may be neglected compared against the coupling into the fiber, i.e., $\kappa_{\rm in}^{(1,2)}\ll \kappa_{\rm ex}^{(1,2)}$. We thus write
\begin{equation}
\kappa=\kappa_{1,2}=\kappa_{\rm ex}^{(1,2)}.
\label{eqn:simplification2}
\end{equation}
We also set
\begin{equation}
h_1=h_2=0,\qquad\hbox{and}\qquad\phi_a=\phi_b=0.
\label{eqn:simplification3}
\end{equation}
With these simplifications we can identify two distinct dynamical behaviors, each with its own distinct features written into the spectrum of spontaneous emission. The distinction arises from the relative phase of the dipole coupling constants, whether the coupling constants have the same phase or their phases differ by $\pi/2$.

\subsection{Coupling between delocalized atomic and photonic excitations: $g_1=g_2=g$}

In the first instance we assume equal dipole coupling constants, $g_1=g_2=g$, which (without loss of generality) we may assume real. We define
\begin{subequations}
\begin{align}
X_\pm&=\frac{1}{\sqrt{2}}\left(\xi_1\pm\xi_2\right),\\
Y_\pm&=\frac{1}{2}\left[\alpha_1+\beta_1\pm(\alpha_2+\beta_2)\right],\\
Z_\pm&=\frac{1}{2}\left[\alpha_1-\beta_1\mp(\alpha_2-\beta_2)\right] ,
\end{align}
\end{subequations}
where for our assumed initial condition, $X_\pm (0)=1/\sqrt{2}$ and $Y_\pm (0)=Z_\pm (0)=0$. The equations of motion, Eqs.~(\ref{eqn:eqns_of_motiona})--(\ref{eqn:eqns_of_motionf}), decouple into a pair of independent sets of three:
\begin{subequations}
\begin{align}
\dot{X}_+&=-\frac{\gamma_A}{2}X_+-i\sqrt{2}g Y_+,\\
\dot{Y}_+&=-2\kappa Y_+-i\sqrt{2}g X_+-\kappa Z_+,\\
\noalign{\vskip4pt}
\dot{Z}_+&=\kappa Y_+,
\end{align}
\end{subequations}
and
\begin{subequations}
\begin{align}
\dot{X}_-&=-\frac{\gamma_A}{2} X_--i\sqrt{2}g Y_-,\\
\dot{Y}_-&=-i\sqrt{2}g X_-+\kappa Z_-,\\
\noalign{\vskip4pt}
\dot{Z}_-&=-2\kappa Z_--\kappa Y_-.
\end{align}
\end{subequations}
The dynamics is dominated by the coupling at rate $\sqrt{2}g$ between $X_\pm$ and $Y_\pm$. Notably this coupling is between {\em delocalised} atomic and photonic excitations---$X_\pm$ and $Y_\pm$ are sums or differences of excitation amplitudes applying to different microtoroids.

From Eqs.~(\ref{eqn:spectrum_fibera}) and (\ref{eqn:spectrum_fiberb}), for the situation considered the spectra of spontaneous emission into the fiber may be written in the form
\begin{subequations}
\begin{align}
T_{\mathrm{fiber},a}(\omega)&=\frac{\kappa}{\pi}\left|\tilde{Y}_+(\omega)+\tilde{Z}_-(\omega)\right|^2\nonumber\\
&=\frac{\kappa}{\pi}\left|\tilde{Y}_+(\omega)-\frac{\kappa}{2\kappa-i\omega}\tilde{Y}_-(\omega)\right|^2,
\label{eqn:spectrum_strong_couplinga}
\end{align}
and
\begin{align}
T_{\mathrm{fiber},b}(\omega)&=\frac{\kappa}{\pi}\left|\tilde{Y}_+(\omega)-\tilde{Z}_-(\omega)\right|^2\nonumber\\
&=\frac{\kappa}{\pi}\left|\tilde{Y}_+(\omega)+\frac{\kappa}{2\kappa-i\omega}\tilde{Y}_-(\omega)\right|^2.
\label{eqn:spectrum_strong_couplingb}
\end{align}
\end{subequations}
As shown by the example in the upper frame of Fig.~\ref{fig:fig2}, these spectra are similar and dominated by vacuum Rabi resonances at frequencies close to $\omega-\omega_A=\pm\sqrt{2}g$. Note that the example takes $\kappa_{\rm i}^{(1,2)}$ smaller than $\kappa_{\rm ex}^{(1,2)}$ but not equal to zero, so spectra are computed from the full set of amplitude equations rather than Eqs.~(\ref{eqn:spectrum_strong_couplinga}) and (\ref{eqn:spectrum_strong_couplingb}). Added to the vacuum Rabi doublet, the weak coupling of $Y_\pm$ to $Z_\pm$ give rise to small features around $\omega-\omega_{\rm A}=0$. For clarity, these are shown on a logarithmic scale in the inset, where the broad feature (width $\sim 2\kappa$) is associated with $Z_-$, while the very narrow feature is associated with $Z_+$ and a system eigenstate
\begin{equation}
\ket{\psi} = \frac{\left[g( -a_1^\dagger + b_1^\dagger + a_2^\dagger - b_2^\dagger) - i\kappa ( \sigma_1^+ + \sigma_2^+ )\right]\ket{G}}{\sqrt{2\kappa^2+4g^2}},
\end{equation}
which is dark for emission into the fiber, i.e., $J_a\ket{\psi}=J_b\ket{\psi}=0$; with no emission into the fiber, spontaneous emission to free space modes occurs with a decay rate $\sim\gamma_{\rm A}(\kappa/2g)^2\ll\gamma_{\rm A}$.

\subsection{Coupling between local atom-photon dressed states: $g_1=ig_2=g$ }
\label{sec:strong_coupling_between_local_dressed_states}

We turn now to dipole coupling constants that differ in phase by $\pi/2$, with $g_1=ig_2=g$, where again we may take $g_1=g$ real without loss of generality. In this case it is instructive to consider the following combinations of excitation amplitudes:
\begin{subequations}
\begin{align}
X_\pm&=\frac{1}{\sqrt{2}}\xi_1\pm\frac{1}{2}(\alpha_1+\beta_1),\\
Y_\pm&=\frac{1}{\sqrt{2}}\xi_2\mp i\frac{1}{2}(\alpha_2-\beta_2),
\end{align}
\end{subequations}
and
\begin{subequations}
\begin{align}
V&=\frac{1}{\sqrt{2}}\left(\alpha_1-\beta_1\right),\\
W&=\frac{1}{\sqrt{2}}\left(\alpha_2+\beta_2\right),
\end{align}
\end{subequations}
where for our chosen initial condition, $X_\pm (0)=1/\sqrt{2}$ and $Y_\pm (0)=V(0)=W(0)=0$. From Eqs.~(\ref{eqn:eqns_of_motiona})--(\ref{eqn:eqns_of_motionf}), it is then possible to derive transparent equations of motion by invoking a rotating-wave approximation with respect to the large frequency $g$. We obtain, with the help of the said approximation,
\begin{subequations}
\begin{align}
\dot{X}_\pm&=-\left[\frac{1}{2}\left(\frac{\gamma_A}{2}+\kappa\right)\pm i\sqrt{2}g\right] X_\pm + i\frac{\kappa}{2}Y_\pm ,\\
\dot{Y}_\pm&=-\left[\frac{1}{2}\left(\frac{\gamma_A}{2}+\kappa\right)\pm i\sqrt{2}g\right] Y_\pm + i\frac{\kappa}{2}X_\pm ,
\end{align}
\end{subequations}
and
\begin{subequations}
\begin{align}
\dot{V}&=-\kappa V+\kappa W,\\
\dot{W}&=-\kappa W-\kappa V.
\end{align}
\end{subequations}
In contrast to the previous case, the equations for $X_\pm$ and $Y_\pm$ now couple {\em local normal-mode} excitations at rate $\kappa/2$; note the normal-mode frequencies $\pm\sqrt2g$ as before. From Eqs.~(\ref{eqn:spectrum_fibera}) and (\ref{eqn:spectrum_fiberb}), the fiber output spectra are given by
\begin{widetext}
\begin{subequations}
\begin{align}
T_{\mathrm{fiber},a}(\omega)&=\frac{\kappa}{4\pi}\left|\left[\tilde{X}_+(\omega)+i\tilde{Y}_+(\omega)\right]-\left[\tilde{X}_-(\omega)+i\tilde{Y}_-(\omega)\right]
+\sqrt{2}\left[\tilde{W}(\omega)+\tilde{V}(\omega)\right]\right|^2,\\
\noalign{\vskip4pt}
T_{\mathrm{fiber},b}(\omega)&=\frac{\kappa}{4\pi}\left|\left[\tilde{X}_+(\omega)-i\tilde{Y}_+(\omega)\right]-\left[\tilde{X}_-(\omega)-i\tilde{Y}_-(\omega)\right]
+\sqrt{2}\left[\tilde{W}(\omega)+\tilde{V}(\omega)\right]\right|^2 ,
\end{align}
\end{subequations}
\end{widetext}

\begin{figure}[t]
\begin{center}
\includegraphics[width=8.5cm]{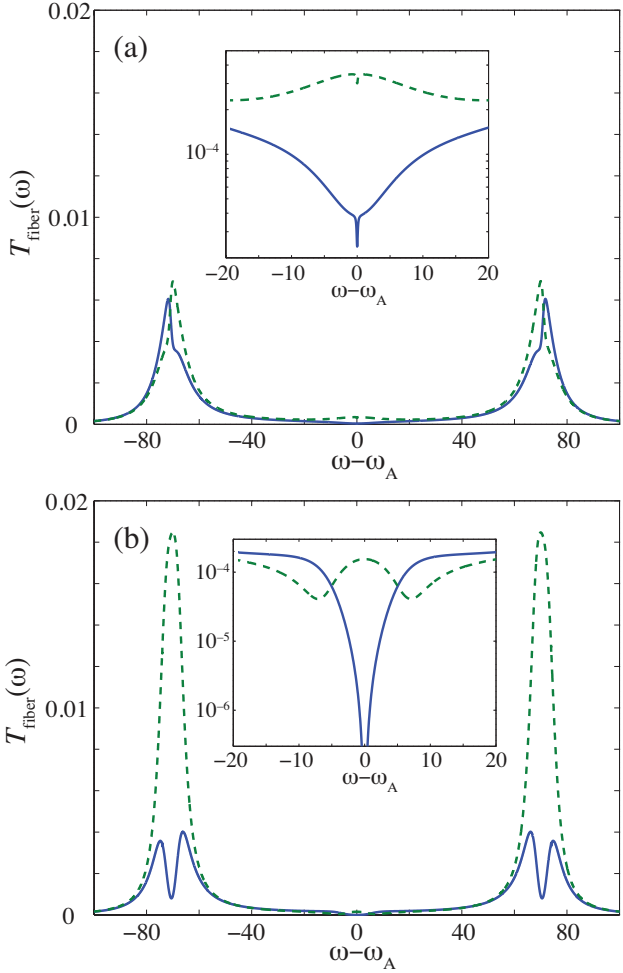}
\caption{Spectrum of spontaneous emission into the fiber from the numerical solution of the full set of coupled amplitude equations; for (in units of $2\pi\cdot$MHz) $\gamma_{\rm A}=5$, $g_1=(g_2, ig_2)=50$ (upper, lower), $\kappa_{\rm ex}^{(1,2)}=5$, $\kappa_{\rm in}^{(1,2)}=0.1$, and all the other parameters zero. The blue solid (green dashed) curves plot the spectrum for the fiber output traveling from left to right (right to left).}
\label{fig:fig2}
\end{center}
\end{figure}

The coupling between normal modes manifests itself prominently in the spectrum of spontaneous emission into the fiber traveling from left to right, where each vacuum Rabi resonance takes the form of a doublet with splitting $\sim\kappa$, as shown in lower frame of Fig.~\ref{fig:fig2}. Emission into the fiber traveling in the opposite direction shows no such splitting. The (small scale) structure around $\omega-\omega_{\rm A}=0$ is determined primarily by the amplitude $W+V$.

\section{Bad-cavity regime}
\label{sec:bad-cavity}

The second regime of special interest is the bad-cavity regime, which corresponds to conditions where the field mode decay rates, $\kappa_1$ and $\kappa_2$, are much larger than the dipole coupling strengths, $g_1$ and $g_2$, and the rate of free-space atomic spontaneous emission $\gamma_A$. In this situation, one may adiabatically eliminate the field modes to arrive at a simpler model describing the atoms alone. We may do this starting from the coupled amplitude equations, Eqs.~(\ref{eqn:eqns_of_motiona})--(\ref{eqn:eqns_of_motionf}), or from the original master equation. We consider both approaches below. We first make the adiabatic elimination in the amplitude equations, from which we recover to two distinct types of effective atomic behavior: superradiant decay in the absence of dipole-dipole interaction, and dipole-dipole interaction without superradiant decay. We then show how this phenomenology emerges from the master equation after adiabatic elimination. The master equation is useful in its own right, as it facilitates our generalization from two to a string of microtoroids in Sec.\ref{sec:many_sites}.

\subsection{Adiabatic elimination of the cavity modes: amplitude equations}

To adiabatic eliminate the field mode amplitudes from Eqs.~(\ref{eqn:eqns_of_motiona})--(\ref{eqn:eqns_of_motionf}), we set the time derivatives of $\alpha_1$, $\beta_1$, $\alpha_2$, and $\beta_2$ to zero and solve the resulting set of algebraic equations to express the field mode amplitudes in terms of $\xi_1$ and $\xi_2$. Substituting the expressions into Eqs.~(\ref{eqn:eqns_of_motiona}) and (\ref{eqn:eqns_of_motiond}) yields the pair of equations:
\begin{subequations}
\begin{align}
\dot\xi_1&=\left(-\frac{\gamma_{\rm A}}{2}-\frac{2|g_1|^2}{\kappa_1+i\Delta_1}\right)\xi_1+\frac{2\sqrt{\kappa_{\rm ex}^{(1)}\kappa_{\rm ex}^{(2)}}g_1^\ast g_2{\rm e}^{i\phi_b}}{(\kappa_1+i\Delta_1)(\kappa_2+i\Delta_2)}\xi_2,\nonumber\\
\label{eqn:bad_cavity_eqn_of_motion1}
\end{align}
and
\begin{align}
\dot\xi_2&=\left( -\frac{\gamma_{\rm A}}{2}-\frac{2|g_2|^2}{\kappa_2+i\Delta_2}\right)\xi_2+\frac{2\sqrt{\kappa_{\rm ex}^{(1)}\kappa_{\rm ex}^{(2)}}g_1^\ast g_2{\rm e}^{i\phi_a}}{(\kappa_1+i\Delta_1)(\kappa_2+i\Delta_2)}\xi_1,\nonumber\\
\label{eqn:bad_cavity_eqn_of_motion2}
\end{align}
\end{subequations}
With the simplifying assumptions of Eqs.~(\ref{eqn:simplification1})--(\ref{eqn:simplification3}), and for the chosen initial condition, $\xi_1(0) = 1$ and $\xi_2(0) = 0$, the solutions are
\begin{subequations}
\begin{align}
\xi_1(t)&=\frac{e^{\lambda_+ t}+e^{\lambda_- t}}2+\frac{|g_1|^2-|g_2|^2}{\kappa+i\Delta}\frac{e^{\lambda_+ t}-e^{\lambda_- t}}p,\nonumber\\
\noalign{\vskip-8pt}\\
\xi_2(t)&=\frac{2\kappa g_1^\ast g_2}{(\kappa+i\Delta )^2}\frac{e^{\lambda_+t}- e^{\lambda_-t}}p,
\end{align}
\end{subequations}
where
\begin{equation}
\lambda_\pm=-\frac{\gamma_{\rm A}}{2}-\frac{|g_1|^2+|g_2|^2}{\kappa+i\Delta}\pm\frac12p,
\end{equation}
with
\begin{equation}
p=\frac{2}{\kappa+i\Delta}\sqrt{\left(|g_1|^2-|g_2|^2\right)^2+\left(\frac{2\kappa g_1^\ast g_2}{\kappa+i\Delta}\right)^2}.
\end{equation}
For dipole coupling constants of the same magnitude, but possibly different phases, i.e., $g_1=e^{-i\theta}g_2=g$ (real), the solutions reduce to the simpler form
\begin{subequations}
\begin{align}
\xi_1(t)&=\frac12\left(e^{\lambda_+t}+e^{\lambda_-t}\right),\\
\xi_2(t)&=\frac12\left(e^{\lambda_+t}-e^{\lambda_-t}\right),
\end{align}
\end{subequations}
with
\begin{equation}
\lambda_\pm=-\frac{\gamma_{\rm A}}{2}-\frac{2g^2}{\kappa +i\Delta}\pm\frac{2\kappa g^2{\rm e}^{i\theta}}{(\kappa +i\Delta )^2},
\label{eqn:eigenvalues}
\end{equation}
or, setting $\Delta=0$,
\begin{equation}
\lambda_\pm=-\frac{\gamma_{\rm A}}{2}-\frac{2g^2}{\kappa}\left(1\mp{\rm e}^{i\theta}\right).
\end{equation}	
This expression for the dynamical eigenvalues reveals the two advertised regimes of control over the field-mediated atom-atom interaction,
first superradiant decay in the absence of dipole-dipole interaction ($\theta=0$), and dipole-dipole interaction without superradiant decay ($\theta=\pi/2$). More generally, from Eq.~(\ref{eqn:eigenvalues}), the detuning can make the selection; for example, dipole-dipole interaction without superradiant decay is also selected with $\theta=0$ and $\Delta=\kappa$. For simplicity, we consider only $\Delta=0$ in what follows.
\subsubsection{Superradiant decay: $\theta =0$}

With both $\Delta$ and $\theta$ set to zero, the dynamical eigenvalues reduce to
\begin{equation}
\lambda_+=-\frac{\gamma_{\rm A}}{2},\qquad\lambda_-=-\frac{\gamma_{\rm A}}{2}-\frac{4g^2}{\kappa}.
\end{equation}	
The term $4g^2/\kappa$ added to the eigenvalue $\lambda_-$ indicates two-atom superradiant decay \cite{Dicke54,Lehmberg70} (see also Sec.~IIIA of \cite{Clemens03}). A single atom coupled to one field mode in the bad cavity limit decays at the cavity-enhanced rate $\gamma_{\rm A}/2+g^2/\kappa$. Spontaneous decay through the \emph{two} output channels of a single microtoroid (coupling to two counterpropagating modes) multiplies the enhanced decay rate by a factor of 2. The second factor of 2, resulting in a $4g^2/\kappa$ enhancement, comes from the collective character of the two-atom cavity-enhanced decay. The eigenvectors corresponding to the eigenvalues $\lambda_\pm$ are
\begin{equation}
\ket{\psi_\pm}=\frac{1}{\sqrt{2}}(\sigma_1^+\pm\sigma_2^+)\ket{G}.
\label{eqn:eigenvectors}
\end{equation}	
These are the familiar subradiant and superradiant states of two-atom collective spontaneous emission.

\begin{figure}[t]
\begin{center}
\includegraphics[width=8.5cm]{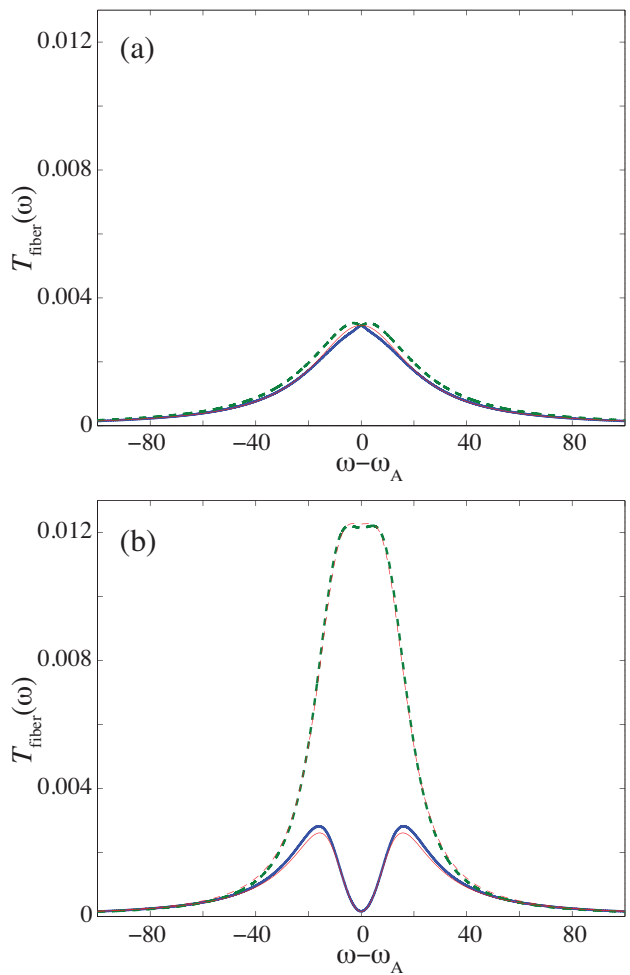}
\caption{Spectra of spontaneous emission into the fiber from the numerical solution of the full set of coupled amplitude equations; for (in units of $2\pi\cdot$MHz) $\gamma_{\rm A}=5$, $g_1=(g_2,-ig_2)=50$ (upper, lower), $\kappa_{\rm ex}^{(1,2)}=500$, $\kappa_{\rm in}^{(1,2)}=0.5$, and all other parameters zero. The blue solid (green dashed) curves plot the spectrum for the fiber output traveling from left to right (right to left). The red solid and dashed lines are plotted from Eq.~(\ref{eqn:spectrum_superradiant}) (upper) and  Eqs.~(\ref{eqn:spectrum_dipole-dipolea}) and (\ref{eqn:spectrum_dipole-dipoleb}) (lower).}
\label{fig:fig3}
\end{center}
\end{figure}

The spectra of spontaneous emission into the fiber are, from Eqs.~(\ref{eqn:spectrum_fibera}) and (\ref{eqn:spectrum_fiberb}) with $\kappa_{\rm ex}^{(1)}=\kappa_{\rm ex}^{(2)}=\kappa$ and $\phi_a=\phi_b=0$,
\begin{subequations}
\begin{align}
T_{\mathrm{fiber},a}(\omega)&=\frac{\kappa}{\pi}|\tilde{\alpha}_1(\omega)+\tilde{\alpha}_2(\omega)|^2,\\
\noalign{\vskip2pt}
T_{\mathrm{fiber},b}(\omega)&=\frac{\kappa}{\pi}|\tilde{\beta}_1(\omega)+\tilde{\beta}_2(\omega)|^2.
\end{align}
\end{subequations}
Since, in a bad cavity with $\theta=0$,
\begin{equation}
\alpha_1+\alpha_2=-(\beta_1+\beta_2)=-i\frac{g}{\kappa}(\xi_1-\xi_2),\nonumber
\end{equation}
both reduce to the Lorentzian
\begin{equation}
T_{\mathrm{fiber},a}(\omega)=T_{\mathrm{fiber},b}(\omega)=\frac{1}{\pi}\frac{g^2}{\kappa}\frac{1}{\lambda_-^2+\omega^2} .
\label{eqn:spectrum_superradiant}
\end{equation}
Hence, spontaneous emission in both directions displays the superradiant linewidth for our assumed initial condition.
The upper frame of Fig.~\ref{fig:fig3} compares the Lorentzian of Eq.~(\ref{eqn:spectrum_superradiant}) with numerical results computed from the full set of coupled amplitude equations. The agreement is very good, and improves with increasing $\kappa$ as expected. Note that the adopted parameters are similar to those of recent experiments with cesium atoms and microtoroidal resonators \cite{Aoki09}.

\subsubsection{Dipole-dipole interaction: $\theta=\pi/2$}
\label{sec:bad_cavity_dipole-dipole_interaction}

With $\Delta$ set to zero and $\theta$ set to $\pi/2$, the dynamical eigenvalues become
\begin{equation}
\lambda_\pm=-\frac{\gamma_{\rm A}}{2}-\frac{2g^2}{\kappa}\pm i\frac{2g^2}{\kappa},
\end{equation}	
with corresponding eigenvectors given by Eq.~(\ref{eqn:eigenvectors}) once again. In this configuration, the cavity-enhanced decay is no longer superradiant, but replaced by a dipole-dipole coupling of magnitude $2|g|^2/\kappa$ between the atoms. With excitation amplitudes satisfying,
\begin{subequations}
\begin{align}
\alpha_1+\alpha_2=i\frac g\kappa(\xi_1+i\xi_2),\qquad\beta_1+\beta_2=-i\frac g\kappa(\xi_1-i\xi_2),\nonumber
\end{align}
\end{subequations}
the spectra of spontaneous emission into the fiber are
\begin{subequations}
\begin{align}
T_{\mathrm{fiber},a}(\omega)&=\frac{1}{4\pi}\frac{g^2}{\kappa}\left|\frac{1-i}{\lambda_++i\omega}-\frac{1+i}{\lambda_-+i\omega}\right|^2,
\label{eqn:spectrum_dipole-dipolea}\\
\noalign{\vskip2pt}
T_{\mathrm{fiber},b}(\omega)&=\frac{1}{4\pi}\frac{g^2}{\kappa}\left|\frac{1+i}{\lambda_++i\omega}-\frac{1-i}{\lambda_-+i\omega}\right|^2.
\label{eqn:spectrum_dipole-dipoleb}
\end{align}
\end{subequations}
These expressions yield different emission spectra in the two output directions, as illustrated in the lower frame of Fig.~\ref{fig:fig3}. For the chosen initial condition, only emission into the fiber traveling from left to right gives rise to a spectrum in the form of a doublet; with a peak-to-peak splitting of $\sim 4g^2/\kappa$, the doublet is a clear indication of dipole-dipole-induced frequency shifts. The spectrum of photon emission in the opposite direction exhibits a single broadened peak---the result of constructive rather than destructive interference of emission from the two collective modes. Note the comparison of the adiabatic elimination against the full set of amplitude equations, where the agreement is again very good.

\subsection{Adiabatic elimination of the cavity modes: master equation}

Finally, we use standard techniques (e.g., Sec.~13.2.1 of \cite{Carmichael07}) to perform the adiabatic elimination in the master equation directly, where, working with the simplifying assumptions of Eqs.~(\ref{eqn:simplification1})--(\ref{eqn:simplification3}), the master equation for the reduced density of the two atoms coupled through the adiabatically eliminated fields is found to be
\begin{widetext}
\begin{align}
\dot{\rho}_{\rm A}=&-i\left[H,\rho_{\rm A}\right]+\left(\frac{\gamma_{\rm A}}2+\frac{2\kappa|g_1^2|}{\kappa^2+\Delta^2}\right)\mkern-2muD[\sigma_1^-] \rho_{\rm A}+\left(\frac{\gamma_{\rm A}}2+\frac{2\kappa|g_2^2|}{\kappa^2+\Delta^2}\right)\mkern-2mu D[\sigma_2^-]\rho_{\rm A}\notag\\
&-2\mathrm{Re}\mkern-2mu\left[g_2g_1^*\frac\kappa{(\kappa+i\Delta)^2}\right]\mkern-3mu\left(2\sigma_1^-\rho_{\rm A}\sigma_2^+-\sigma_2^+\sigma_1^-\rho_{\rm A}-\rho_{\rm A} \sigma_2^+ \sigma_1^-+\mathrm{H.c.}\right),
\label{eqn:master_equation_bad_cavity}
\end{align}
with Hamiltonian
\begin{equation}
H=-\frac{2\Delta}{\kappa^2+\Delta^2}(|g_1|^2\sigma_1^+\sigma_1^-+|g_2|^2\sigma_2^+\sigma_2^-)
-2\mathrm{Im}\mkern-2mu\left[g_2g_1^*\frac\kappa{(\kappa+i\Delta)^2}\right]\mkern-2mu(\sigma_1^+\sigma_2^-+\sigma_2^+\sigma_1^-),
\label{eqn:hamiltonian_bad_cavity}
\end{equation}	
\end{widetext}
The collective terms in Eqs.~(\ref{eqn:master_equation_bad_cavity}) and (\ref{eqn:hamiltonian_bad_cavity}) [also (\ref{eqn:master_equation_N_atoms}) and (\ref{eqn:hamiltonian_N_atoms})] generalize the master equation for atoms interacting through a one-dimensional waveguide; in the latter case, the argument of the complex number $g_sg_1^*\kappa/(\kappa+i\Delta)^2$ is replaced by $ik_A|z_1-z_2|$, with $k_A$ the wavenumber at the atomic resonance frequency, and $z_1$ and $z_2$ the atomic positions \cite{Chang12,Lalumiere13,vanLoo13}. From the above general equation, master equations for collective decay without (coherent) dipole-dipole interaction and  dipole-dipole interaction without collective decay are recovered as follows.
\subsubsection{Superradiant decay: $\theta=0$}

Setting both $\Delta$ and $\theta$ to zero, with $g_1=g_2=g$ (real), Eq.~(\ref{eqn:master_equation_bad_cavity}) simplifies as 		
\begin{equation}
\dot{\rho}_{\rm A}=\frac{2g^2}{\kappa}D[J^-]\rho_{\rm A}+\frac{\gamma_{\rm A}}{2}\left(D[\sigma_1^-]\rho_{\rm A}+D[\sigma_2^-]\rho_{\rm A}\right),
\end{equation}
where
\begin{equation}
J^{\pm}=\sigma_1^{\pm}-\sigma_2^{\pm}.
\label{eqn:collective_jump_operators}
\end{equation}
The first term on the right-hand side describes two-atom superradiant decay. Note that for the dynamical eigenvectors of Eq.~(\ref{eqn:eigenvectors}),
\begin{equation}
J^-\ket{\psi_+}=0,\qquad J^-\ket{\psi_-}=\sqrt{2}\ket{G} .
\end{equation}
Note also that the particular form of the collective jump operators in Eq.~(\ref{eqn:collective_jump_operators})---with a minus sign---follows from choosing $g_1$ and $g_2$ to have the same rather than opposite phase (and the choice $\phi_a=\phi_b=0$).

\subsubsection{Dipole-dipole interaction: $\theta=\pi/2$}

Alternatively, setting $\Delta$ to zero and $\theta$ to $\pi/2$, Eq.~(\ref{eqn:master_equation_bad_cavity}) simplifies as
\begin{align}
\dot{\rho}_{\rm A}=&\,\,i\frac{2g^2}{\kappa}\mkern-2mu \left[ \sigma_1^+\sigma_2^-+\sigma_2^+\sigma_1^-,\rho_{\rm A}\right]\nonumber\\
&+\left(\frac{\gamma_{\rm A}}2+\frac{2g^2}{\kappa}\right)\mkern-2mu(D[\sigma_1^-]\rho_{\rm A}+D[\sigma_2^-] \rho_{\rm A}).
\end{align}
This master equation describes cavity-enhanced, single-atom spontaneous emission in the presence of a cavity-mediated dipole-dipole interaction between the atoms.

\section{Generalization to many sites}
\label{sec:many_sites}

\subsection{Master equation}

The master equation in the bad-cavity regime is readily generalized from two to $N$ microtoroid systems coupled through a single optical fiber. Working again with the simplifying assumptions of Eqs.~(\ref{eqn:simplification1})--(\ref{eqn:simplification3}), Eqs.~(\ref{eqn:master_equation_bad_cavity}) and (\ref{eqn:hamiltonian_bad_cavity}) generalize as
\begin{widetext}
\begin{align}
\dot{\rho}_{\rm A}=&-i[H,\rho_{\rm A}]+\sum_{k=1}^N\left(\frac{\gamma_{\rm A}}2+\frac{2|g_k|^2\kappa}{\kappa^2+\Delta^2}\right)D[\sigma_k^-]\rho_{\rm A}\nonumber\\
&-\sum_{k=1}^N\sum_{l=k+1}^N 2{\rm Re}\mkern-2mu\left[g_lg_k^*\frac\kappa{(\kappa+i\Delta)^2}\eta^{l-k-1}\right]\mkern-2mu\left( 2\sigma_l^-\rho_{\rm A}\sigma_k^+ - \sigma_k^+\sigma_l^-\rho_{\rm A} - \rho_{\rm A}\sigma_k^+\sigma_l^- + \mathrm{H.c.}\right),
\label{eqn:master_equation_N_atoms}
\end{align}
with Hamiltonian
\begin{equation}
H=- \frac{2\Delta}{\kappa^2+\Delta^2}\sum_k |g_k|^2\sigma_k^+\sigma_k^-- \sum_{k=1}^N\sum_{l=k+1}^N 2{\rm Im}\mkern-2mu\left[g_lg_k^*\frac{2\kappa}{(\kappa+i\Delta)^2}  \eta^{l-k-1}\right]\mkern-2mu(\sigma_k^+\sigma_l^- + \sigma_l^+\sigma_k^-),
\label{eqn:hamiltonian_N_atoms}
\end{equation}
\end{widetext}
where
\begin{equation}
\eta=-\frac{\kappa-i\Delta}{\kappa+i\Delta}.
\end{equation}
As in the two-atom case, we readily identify interesting limiting cases for special choices of the dipole coupling strengths. We focus on the natural generalizations of the examples discussed in Secs.~\ref{sec:strong-coupling} and \ref{sec:bad-cavity}.

\subsection{Special cases}

First, setting $\Delta$ to zero with $g_1=g_2=\cdots=g$ (real), Eq.~(\ref{eqn:master_equation_N_atoms}) simplifies to the master equation for $N$-atom cavity-mediated superradiant decay:
\begin{equation}
\dot{\rho}_{\rm A}=\frac{2g^2}\kappa D[J^-]\rho_{\rm A}+\sum_{k=1}^N\frac{\gamma_{\rm A}}2D[\sigma_k^-]\rho_{\rm A},
\label{eqn:master_equation_N_atoms_special_case1}
\end{equation}
with
\begin{equation}
J^-=\sigma_1^--\sigma_2^-+\cdots+(-1)^{N+1}\sigma_N^-.
\end{equation}
The second example from Secs.~\ref{sec:strong-coupling} and \ref{sec:bad-cavity}, with a $\pi/2$ phase difference between $g_1$ and $g_2$, does not generalize in quite such an obvious way. We generalize to alternating real and imaginary coupling strengths: $g_1=g_3=\cdots=-ig_4=-ig_6\cdots=g$ (real). This configuration no longer yields dipole-dipole interaction without superradiant decay, but rather a combination of both, where each pair interaction shows a strong dependence on the relative positions of the atoms in the pair. The master equation may be written in the form
\begin{align}
\dot{\rho}_{\rm A}=&-i[H,\rho_{\rm A}] + \sum_{k=1}^N\frac{\gamma_{\rm A}}{2}D[\sigma_k^-]\rho_{\rm A}\nonumber\\
&+\frac{2g^2}{\kappa}(D[J_{\rm odd}^-]\rho_{\rm A}+D[J_{\rm even}^-]\rho_{\rm A}),
\label{eqn:master_equation_N_atoms_special_case2}
\end{align}
with Hamiltonian
\begin{equation}
H=\frac{2g^2}\kappa\sum_{k=1}^N(-1)^k\mkern-12mu\sum_{l-k=1,3,\ldots}(\sigma_l^+\sigma_k^-+\sigma_k^+\sigma_l^-),
\end{equation}
and
\begin{subequations}
\begin{align}
J_{\rm odd}^-&=\sigma_1^-+\sigma_3^-+\sigma_5^-+\cdots,\\
J_{\rm even}^-&=\sigma_2^-+\sigma_4^-+\sigma_6^-+\cdots.
\end{align}
\end{subequations}
Thus, each atom participates in superradiant decay with atoms located an {\em even} number of sites away, and, at the same time, dipole-dipole interaction with atoms located an {\em odd} number of sites away.

\subsection{$N=3$}

As an illustration, we plot spectra for $N=3$ in Figs.~4 and 5. Results are computed from the full set of coupled amplitude equations---i.e., the extension of Eqs.~(\ref{eqn:eqns_of_motiona})--(\ref{eqn:eqns_of_motionf}) to many sites---without the adiabatic elimination. The parameters, however, are chosen to approximate the adiabatic regime, and Eqs.~(\ref{eqn:master_equation_N_atoms_special_case1}) and (\ref{eqn:master_equation_N_atoms_special_case2}) provides clear insight into the features observed.

The spectra of Fig.~6 correspond to conditions where superradiant decay is predicted. They are dominated by a broad single peak of halfwidth $\sim N 2g^2/\kappa=6g^2/\kappa$, with an additional small feature associated with single-atom spontaneous emission at rate $\gamma_{\rm A}$. Notably, there is a dependence on the position of the initially excited atom, i.e., whether it is at the center of the group of three or at one end: the spectra depend on the direction of propagation when the excited atom is at one end, but are independent of direction---by symmetry---when the center atom is excited. Differences in the former case are seen from our numerical solution to become smaller the larger $\kappa$ becomes relative to the dipole coupling $g$.

\begin{figure}[t]
\begin{center}
\includegraphics[width=8cm]{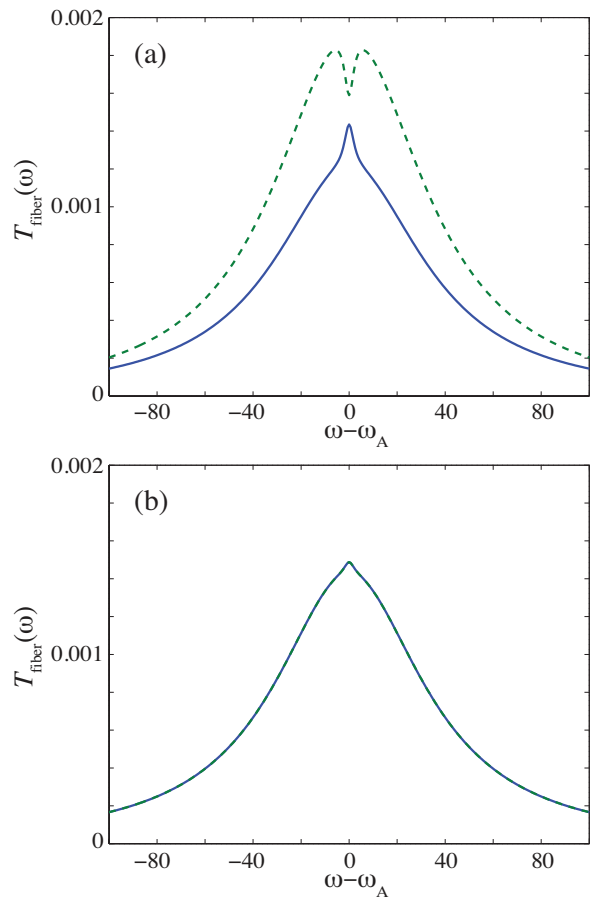}
\caption{Spectra of spontaneous emission into the fiber in the bad cavity regime for three atoms. Curves plotted from the numerical solution of the full set of coupled amplitude equations; for (in units of $2\pi\cdot$MHz) $\gamma_{\rm A}=5$, $g_1=g_2=g_3=50$, $\kappa_{\rm ex}^{(1,2,3)}=500$, $\kappa_{\rm in}^{(1,2,3)}=0.5$, and all other parameters zero. The blue solid (green dashed) curves plot the spectrum for the fiber output traveling from left to right (right to left). The initial state is $\sigma_1^+\ket{G}$ (upper) and $\sigma_2^+\ket{G}$ (lower).
}
\label{fig:fig4}
\end{center}
\end{figure}

\begin{figure}[t]
\begin{center}
\includegraphics[width=8cm]{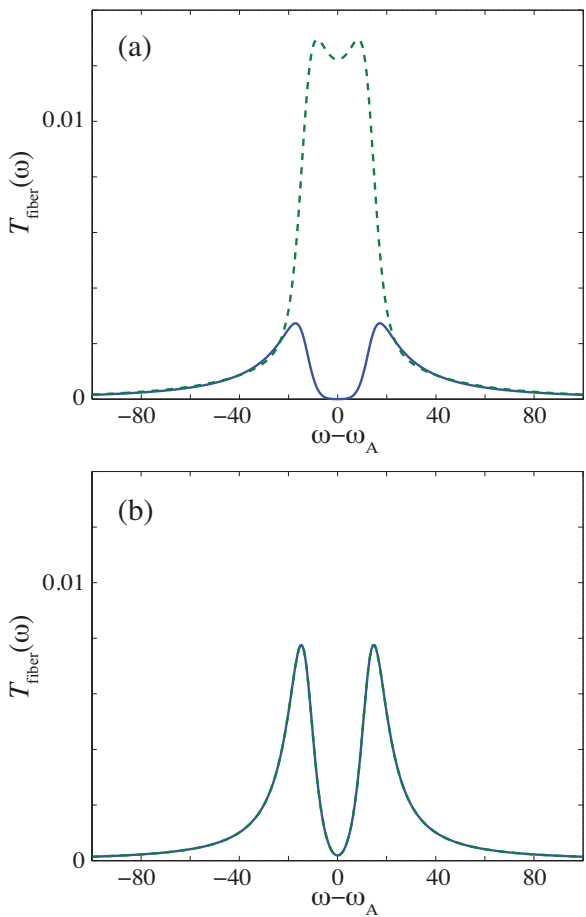}
\caption{As in Fig.~\ref{fig:fig4} but with $g_1=-ig_2=g_3=50$.}
\label{fig:fig5}
\end{center}
\end{figure}

The spectra of Fig.~\ref{fig:fig5} take $g_1=-ig_2=g_3=g$, where, for three atoms in this configuration, the master equation in the adiabatic approximation, Eq.~(\ref{eqn:master_equation_N_atoms_special_case2}), is
\begin{align}
\dot{\rho}_{\rm A}=&-i[H,\rho_{\rm A}]+\frac{\gamma_{\rm A}}{2}(D[\sigma_1^-]+D[\sigma_2^-] + D[\sigma_3^-])\rho_{\rm A},\nonumber\\
&+\frac{2g^2}{\kappa}D[\sigma_1^-+\sigma_3^-]\rho_{\rm A}+\frac{2g^2}{\kappa}D[\sigma_2^-]\rho_{\rm A},
\end{align}
with dipole-dipole Hamiltonian
\begin{equation}
H=\frac{2g^2}{\kappa}\left( -\sigma^+_2\sigma^-_1-\sigma^+_1\sigma^-_2+\sigma^+_3\sigma^-_2+\sigma^+_2\sigma^-_3\right).
\label{eqn:hamiltonian_3_atoms_special_case2}
\end{equation}
The atoms on the ends participate in cavity-mediated superradiant decay, while the center atom exhibits only cavity-enhanced single-atom spontaneous emission. While not participating in the superradiant decay, the center atom is nonetheless coupled to its neighbors through cavity-mediated dipole-dipole interactions.

Diagonalizing dipole-dipole Hamiltonian (\ref{eqn:hamiltonian_3_atoms_special_case2}) yields the pair of eigenkets
\begin{subequations}
\begin{equation}
|e_\pm\rangle=\frac{1}{2}\left(\sigma_1^+\mp\sqrt{2}\sigma_2^+-\sigma_3^+\right)\ket{G},
\end{equation}
with eigenvalues $e_\pm=\pm2\sqrt{2}g^2/\kappa$, and the eigenket
\begin{equation}
|e_0\rangle=\frac{1}{\sqrt{2}}\left(\sigma_1^++\sigma_3^+\right)\ket{G},
\end{equation}
\end{subequations}
with eigenvalue $e_0=0$. Features appear at corresponding frequencies in the spectra of Fig.~\ref{fig:fig5}. Note the action of the jump operator $J_{\rm odd}^-=\sigma_1^-+\sigma_3^-$ on the three eigenkets: $J_{\rm odd}^-|e_+\rangle=J_{\rm odd}^-|e_-\rangle=0$ and $J_{\rm odd}^-|e_0\rangle=\sqrt{2}\ket{G}$. Thus, with initial state $\sigma_1^+\ket{G}=(|e_+\rangle+|e_-\rangle+\sqrt{2}|e_0\rangle)/2$, one observes contributions to the emission from all three eigenstates, albeit in different combinations depending on the emission direction. On the other hand, for initial state $\sigma_2^+\ket{G}=(1/\sqrt{2})\left(|e_+\rangle-|e_-\rangle \right)$, superradiant decay is absent from the dynamics and the spectrum takes a straightforward double-peaked form as a result of the dipole-dipole interaction between neighboring atoms.

\section{Conclusion}
\label{sec:conclusion}

We have taken a first step towards characterising and manipulating the emission properties of two-level atoms through their coupling, via evanescent fields, to a network of fiber-linked microtoroidal resonators. Using the theory of two-way cascaded systems, we computed the spectrum of spontaneous emission into the fiber outputs for a pair of atom-microtoroid systems and illucidated its dependence on the relative phase of the atom-cavity coupling: under strong-coupling conditions, the decay is dominated by an interplay of either delocalized or local excitations, depending on relative phase, while under bad-cavity conditions (microtoroid and fiber fields adiabatically eliminated), the atoms exhibit either collective spontaneous emission with no dipole-dipole interaction, or a (coherent) dipole-dipole interaction and single-atom, cavity-enhanced spontaneous emission. In the latter case, we recover behavior expected for atoms coupled through a one-dimensional waveguide from our two-way cascaded system model.
 
We extended our results under bad-cavity conditions to many two-way cascaded atom-microtoroid systems. While the conditions leading to collective spontaneous emission with no dipole-dipole interaction carry over, a more complex combination of collective decay---involving atoms separated by an even number of sites---and dipole-dipole interaction---atoms separated by an odd number of sites---is seen in the second case.

An externally driven network raises interesting issues untouched in this work; it may attain a non-trivial steady state, e.g., one showing significant entanglement between the atoms \cite{Gupta10,Ramos14}. Preliminary results on the driven version of our two-way cascaded system are presented in \cite{Zeeb09}, though this is a direction wide open for further study.

\section*{ACKNOWLEDGMENTS}
This work was supported by the Marsden fund of the Royal Society of New Zealand.



\end{document}